\documentclass[aps,prc,twocolumn,showpacs,nofootinbib,floatfix,superscriptaddress]{revtex4-1}
\usepackage{graphicx}
\usepackage{amsfonts}
\usepackage{amsmath}
\usepackage{hyperref}
\usepackage{float}
\usepackage{verbatim} 

\begin{document}

\title{Transition From Ideal To Viscous Mach Cones In A Kinetic Transport Approach}

\author{I.\ Bouras}
\affiliation{Institut f\"ur Theoretische Physik,
Johann Wolfgang Goethe-Universit\"at,
Max-von-Laue-Str.\ 1, D-60438 Frankfurt am Main, Germany}

\author{A.\ El}
\affiliation{Institut f\"ur Theoretische Physik,
Johann Wolfgang Goethe-Universit\"at,
Max-von-Laue-Str.\ 1, D-60438 Frankfurt am Main, Germany}

\author{O.\ Fochler}
\affiliation{Institut f\"ur Theoretische Physik,
Johann Wolfgang Goethe-Universit\"at,
Max-von-Laue-Str.\ 1, D-60438 Frankfurt am Main, Germany}

\author{H.\ Niemi}
\affiliation{Department of Physics, P.O. Box 35 (YFL), FI-40014 University of
Jyv\"askyl\"a, Finland}

\author{Z.\ Xu}
\affiliation{Department of Physics, Tsinghua University, Beijing 100084, China}

\author{C.\ Greiner}
\affiliation{Institut f\"ur Theoretische Physik,
Johann Wolfgang Goethe-Universit\"at,
Max-von-Laue-Str.\ 1, D-60438 Frankfurt am Main, Germany}

\begin{abstract}

Using a microscopic transport model we investigate
the evolution of conical structures originating from the supersonic
projectile moving through the hot matter of ultrarelativistic particles.
Using different scenarios for the interaction between projectile and matter,
and different transport properties of the matter, we study the formation 
and structure of Mach cones. Especially, a dependence of the Mach cone 
angle on the details and rate of the energy deposition from projectile 
to the matter is investigated. Furthermore, the two-particle correlations extracted 
from the numerical calculations are compared to an analytical approximation.
We find that the propagation of a high energetic particle through the matter 
does not lead to the appearance of a double peak structure as observed in
the ultrarelativistic heavy-ion collision experiments. The reason is the 
strongly forward-peaked energy and momentum deposition in the head shock
region. In addition, by adjusting the cross section we investigate the
influence of the viscosity to the structure of Mach cones. A clear and
unavoidable smearing of the profile depending on a finite ratio of shear
viscosity to entropy density is clearly visible.

\end{abstract}


\date{\today}

\maketitle

\section{Introduction}

Results from the relativistic Heavy Ion Collider (RHIC) \cite{brahms}
and recently from the Large Hadron Collider (LHC) \cite{Aamodt:2010pa}
indicate the formation of a new state of matter, the quark-gluon plasma
(QGP). The large value of the measured elliptic flow coefficient $v_{\rm 2}$
indicates the nearly perfect fluid behavior of the QGP \cite{Adler:2003kt}.
This is confirmed by recent calculations of viscous hydrodynamics
\cite{Luzum:2008cw} and microscopic transport calculations \cite{Xu:2007jv}
with a shear viscosity over entropy density ratio $\eta/s = 0.1 -0.2$,
which is close to the conjectured lower bound $\eta/s = 1/4\pi$ from a
correspondence between conformal field theory and string theory in an
Anti-de-Sitter space \cite{Kovtun:2004de}.

Highly energetic partons propagating through
the hot and dense QGP rapidly lose their energy and momentum as the energy
is deposited in the medium. This phenomenon is known as
jet-quenching \cite{Adams:2003kv,Gyulassy:2004zy}, whereas
its exact mechanism is still to be fully understood. Furthermore, recent 
measurements of two- and three-particle correlations in heavy-ion 
collisions (HIC) show a complete suppression of the away-side jet, 
whereas for lower $p_T$ a double peak structure is observed 
in the two-particle correlation function \cite{Wang:2004kfa}. 
For a while one possible and promising origin of these structures
was assumed to be the interaction of fast partons with the soft matter
which generates collective motion of the medium in form of Mach cones.
\cite{Stoecker:2004qu,Bouras:2010nt}.
In contrast, recent studies of triangular flow from initial
fluctuations \cite{Ma:2010dv} show a more satisfactory explanation
for the appearance of the double peak structure.

The recent idea is that both Mach cones and triangular flow from
initial fluctuations exist in heavy-ion collisions, but it is difficult
to separate their effects and the two-particle correlations do not
seem to be a good observable for this purpose. In the present study
we claim that even if there are no effects from
initial stage fluctuations, a double-peak structure in two-particle
correlations cannot be expected from the energy-momentum deposition
by a jet into the medium. This will be not only due to the viscous
effects, but also due to the details of the "Mach cone"-like
structure, which was found in a similar form in ideal fluid
\cite{Betz:2008ka} and AdS/CFT\cite{Noronha:2008un} studies.

For this purpose we investigate the propagation and formation of
Mach cones in the microscopic transport model BAMPS
(Boltzmann Approach of MultiParton Scatterings) \cite{Xu:2004mz}
in the limit of vanishing mass and very small shear viscosity over
entropy density ratio $\eta/s$ of the matter. Two different scenarios
for the jet are used and the dependence of the Mach cone angle on the
details of energy deposition is discussed. A simple analytic relation for
the expected particle distribution in the Mach cone wings is derived
and is compared it to numerical results extracted from BAMPS. In addition,
by adjusting $\eta/s$, the influence of the viscosity on the profile of
the Mach cone and the corresponding two-particle correlation is explored
for the first time. In this work the units are $\hbar = c = k = 1$.
The metric tensor is $g^{\mu \nu} = \textrm{diag}(1,-1,-1,-1)$.

\section{Shock Waves and Mach cones}
\label{sec:shocksAndMachCones}

Shock waves are phenomena which have their origin in
the collective motion of matter. In the limit of a perfect fluid
with no viscosity, a signal caused by a weak perturbation 
propagates with the speed of sound $c_{\rm s} = \sqrt{dp/de}$, which depends
on the equation of state (EOS) of the medium. Here, $p$ is the
equilibrium pressure and $e$ is the energy density in the local 
rest frame (LRF). A larger perturbation results in a shock wave 
propagating faster than the speed of sound. In a simplified 
one-dimensional setup shock waves have already been studied for the 
perfect fluid limit \cite{Schneider:1993gd}. Furthermore, the viscous
solutions have been investigated in Refs.~\cite{Bouras:2009nn,Molnar:2008fv}, 
demonstrating that the shock profile is smeared out when viscosity is large.
It was also found that a clear observation of the shock within
the short time available in heavy-ion collisions requires a
small viscosity. The information taken from
these studies can be transferred to the investigation
of conical shock structures like Mach cones, which is the
main subject of this Letter.

First, we consider a weak perturbation moving with the speed
of light, i.e. $v_{\rm source} = 1$, through the medium of a
perfect fluid. For simplification we use a massless relativistic
gas with $e = 3p$ and $c_{\rm s} = 1/\sqrt{3}$. The perturbation
generates waves propagating through the medium with the speed of sound
$c_{\rm s}$. In this case the propagating modes are called sound waves.
If the perturbation moves faster than the speed of sound the
created sound waves accumulate on a cone \cite{Landau}. In the following
we refer to the surface of this cone as the shock front. 

The resulting emission angle of this shock front relative to the direction 
of the projectile is given by the weak perturbation Mach angle
$\alpha_{\rm w} = \arccos (c_{\rm s} / v_{\rm source}) = \arccos (1 / \sqrt{ 3 }) = 54,73^\circ$.
It is important to know that in nature perturbations are not sufficiently
small. In this case, shock waves instead of sound waves are generated and
due to different propagation velocities of these waves, we expect a change
of the Mach angle \cite{Rischke:1990jy}. We can generalize the introduced
Mach angle to the case of stronger perturbations:
\begin{equation}
\label{eq:IdealMachAngle}
\alpha = \arccos (v_{ \rm shock} / v_{\rm source} ) \, .
\end{equation}
We require here $v_{\rm source} > v_{\mathrm{shock}}$,
where $v_{\rm shock}$ is the velocity of the shock front
propagating through the medium.
The velocity of the shock front depends on the pressure
(energy density) on the cone $p_{ \rm cone}$ ($e_{ \rm cone}$)
and the medium itself $p_{ \rm med}$ ($e_{ \rm med}$)
\cite{Schneider:1993gd}:
\begin{equation}
\label{eq:v_shock}
v_{\rm{shock}} = \left [ \frac{(p_{ \rm med} - p_{ \rm cone})(e_{ \rm cone} + p_{ \rm med})}
{(e_{ \rm med} - e_{ \rm cone})(e_{ \rm med} + p_{ \rm cone})} \right]^{1/2} \, .
\end{equation}
Eq.~\eqref{eq:v_shock} has the following limits: If
$p_{ \rm cone} \gg p_{ \rm med}$ we obtain $v_{\rm{shock}} = 1$. If
$p_{ \rm cone} \approx p_{ \rm med}$, i.e. the perturbation is very weak,
we get the expected limit of the speed of sound
$v_{\rm{shock}} \approx c_s$. In the latter case
Eq.~\eqref{eq:IdealMachAngle} becomes $\alpha_w$,
as expected. The collective velocity of matter in the
shock wave (Mach cone wing), which is different from
the signal propagation velocity \eqref{eq:v_shock}, can be
calculated via
\begin{equation}
\label{eq:v_coll}
v_{\rm{coll}} = \left [ \frac{(p_{ \rm cone} - p_{ \rm med})(e_{ \rm cone} - e_{ \rm med})}
{(e_{ \rm med} + p_{ \rm cone})(e_{ \rm cone} + p_{ \rm med})} \right]^{1/2} \, .
\end{equation}
In the case of a very weak perturbation the collective
velocity of matter vanishes, $v_{\rm{coll}} \approx 0$,
whereas for stronger perturbations $v_{\rm{coll}}$
can increase up to the speed of light.

In Sec.~\ref{sec:BAMPS} we discuss the numerical results
from BAMPS and expect a clear dependence of the observed
Mach angle on the strength of the perturbation according to
Eq.~\eqref{eq:IdealMachAngle}, but due to non-linear
effects Eq.~\eqref{eq:IdealMachAngle} is merely a good
approximation.

%
\section{Particle Momentum Distribution in the Shock Front}
\label{sec:distributionTwoParticle}
%

In order to understand the origin of the double peak structure
induced by "Mach cone"-like structures, which will be discussed in
Sec.~\ref{sec:BAMPS}, we derive a simple model of particle
emission from the shock front of a Mach cone in a 2-dimensional
$xy$-plane. We assume two sources
modeling the two wings of a Mach cone with a constant temperature
$T$ and collective four-velocity $u^{\mu} = \gamma (1,\vec{v})$, 
where $\gamma = 1 / \sqrt{1 - v^2}$ is the Lorentz gamma factor. 
Each source consists of massless particles according to the thermal distribution
$f(\boldsymbol{x},\boldsymbol{p}) = exp({-u_{\mu}p^{\mu}/T})$,
where $p^{\mu} = (E,\vec{p})$ is the particle four-momentum. 
Choosing the $x$-axis to be the symmetry axis of the cone, which is 
simultaneously the propagation direction of the jet, we can write
$u_{\pm}^{\mu} = \gamma(1,v \cos{\alpha}, \pm v \sin{\alpha},0)$.
The $\pm$ corresponds to each wing of the cone. We identify
$v = v_{\rm coll}$ with Eq.~\eqref{eq:v_coll} as the collective
velocity of the matter in the shock wave and $\alpha$ is the
Mach angle defined in Eq.~\eqref{eq:IdealMachAngle}. Using the same coordinate
system we write for the four-momentum vector
$p^{\mu} =p(1,\cos{\phi} \sin{\theta},\sin{\phi} \sin{\theta},\cos{\theta})$.
$\phi$ is the azimuthal angle in the $xy$ plane and
$\theta$ is the polar angle with the $z$-axis.

The distribution function is defined as
$dN (2\pi)^3/(dV d^3p) = f(\boldsymbol{x},\boldsymbol{p})$,
where $dV \, d^3p / (2\pi)^3$ is the phase space volume element.
We are interested in the particle distribution
$dN/(N  d\phi)$ which can be calculated as an integral
over the thermal distribution in a certain volume $V$ on
the Mach cone surface. We use
$d^3p = p^2 \, dp \, d\phi \, d(\cos \theta)$ and write
\begin{equation}
\label{eq:particleMultiCone_general}
\frac{{\rm d}N}{N {\rm d}\phi} = \frac{V}{N (2 \pi)^3} \int \limits_{0}^{\pi} {\rm d}\theta \sin{\theta} \int \limits_{0}^{\infty} p^2 \left( e^{-\frac{ u_{+}^{\mu} p_{\mu} }{T} } + e^{-\frac{u_{-}^{\mu} p_{\mu}}{T} }  \right) {\rm d}p
\text{.}
\end{equation}
We obtain $N = 8\pi\gamma T^3 V$ by integrating
$dN/(dV d^3p) = f(\boldsymbol{x},\boldsymbol{p})$
over the entire phase space volume. After the integration
of \eqref{eq:particleMultiCone_general} we obtain
\begin{equation}
\label{eq:particleMultiCone}
\frac{dN}{N d\phi} = \frac{1}{ 8 \pi \gamma^4} \displaystyle\sum\limits_{i=1}^2  \left[ \frac{2+b_i^2}{(1-b_i^2)^2} + \frac{3b_i}{(1-b_i^2)^{5/2}}A \right] \, ,
\end{equation}
where $A = \pi/2 + \arctan(b_i/\sqrt{1-b_i^2})$, $b_1 = v \cos (\alpha - \phi)$
and $b_2 = v \cos (\alpha + \phi)$.

The most important result taken from Eq.~\eqref{eq:particleMultiCone} is
the non-existent double peak structure for small $v_{\rm coll}$, which is
against all expectations resulting from the naive picture of a Mach cone.
We will discuss this point in more details in Sec.~\ref{sec:BAMPS} by
comparing Eq.~\eqref{eq:particleMultiCone} to the numerical results.
Furthermore, the particle distribution is equivalent to the two-particle
correlation, since the angle $\phi$ is always correlated to the direction
of the source, which serves as a "trigger" particle.

%
\section{Transition from Ideal to Viscous Mach Cones in BAMPS}
\label{sec:BAMPS}
%

In the following we study the evolution of "Mach cone"-like
structures with different scenarios of the jet-medium interaction
by using the parton cascade BAMPS \cite{Xu:2004mz} -- a
microscopic transport model which solves the Boltzmann equation
$p^{\mu} \partial_{\mu} f(x,p) = C \left[f(x,p) \right]$
for on-shell particles based on stochastic interpretation
of transition rates. As was demonstrated in previous works,
BAMPS is able to
explore a large variety of hydrodynamic phenomena and
provides a reliable benchmark for hydrodynamic models
\cite{Bouras:2009nn,El:2011cp}.
The advantage of BAMPS is its ability to handle arbitrary
large gradients for any choice of viscosity. Thus it is
possible to investigate the complete transition from ideal
to viscous behavior.

In this study we focus on investigation of Mach cone evolution
in absence of any other effects - i.e. we neglect such effects
as initial fluctuations or expansion, which are however relevant
in heavy-ion collisions. For this purpose, the space-time
evolution of particles is performed in a static box. We
initialize a static uniform medium of massless Boltzmann
particles with $T_{\rm med} = 400$ MeV, which corresponds
to a LRF energy density $e_{\rm med} = 16.28 \, \rm{GeV/fm^3}$.
For simplification, we consider only binary collisions with
an isotropic cross section, i.e. a cross section with an
isotropic distribution of the collision angle. Furthermore,
we keep the mean free path $\lambda_{\rm mfp}$ of the medium
particles constant in all spatial cells by adjusting the
cross section according to $\sigma = 1 / (n\lambda_{\rm mfp})$,
where $n$ is the particle density. The related shear viscosity
for isotropic binary collisions is given by
$\eta = 0.4\,e \, \lambda_{\rm mfp}$ \cite{deGroot}.
Collisions of particles against box boundaries in $x$ and
$y$ direction are realized as elastic collisions off a wall;
in $z$-direction we use periodic boundary conditions. This
reduces the problem to two dimensions and therefore decreases
the numerical expenses.

We introduce two different sources to investigate the evolution
of "Mach cone"-like structures. In the so called pure energy
deposition scenario (PED) \cite{Betz:2008ka}
the source propagates and emits particles according to the
thermal distribution $f(x,p) = exp(-E/T)$, so that the energy
deposition is isotropic in the LRF of the source. In this
scenario on average only energy is deposited to the medium,
but no net-momentum. In the second scenario, referred to as
JET, a highly energetic massless particle (jet) has only
momentum in $x$-direction, i.e. $p_{\rm x} = E_{ \rm jet}$.
The jet propagates and deposits energy to the medium due to
collisions with particles. After each collision, the momentum of
the jet is reset to its initial value. The jet-medium cross section
is adjusted in such a way that we obtain a specific energy
deposition rate. Using this scenario a constant energy and
momentum deposition rate is achieved. For both scenarios the
sources are initialized at $t = 0$ fm/c at the position
$x = - 0.1$ fm and propagate in $x$-direction with
$v_{\rm source} = 1$, i.e. with the speed of light.
We note the JET scenario is a simplified model of a jet
in heavy-ion physics, whereas the
PED scenario vaguely resembles the hot spots
studied in \cite{Takahashi:2009na}, but in the form implemented
here there is no correspondence to heavy-ion collisions. 
We expect clear differences between these two
scenarios concerning the evolution of the entire system, but
also concerning the final distribution of the particles.

\subsection{Effect of energy deposition rate}

\begin{figure*}[ht]
\centering 
\includegraphics[width=\textwidth]{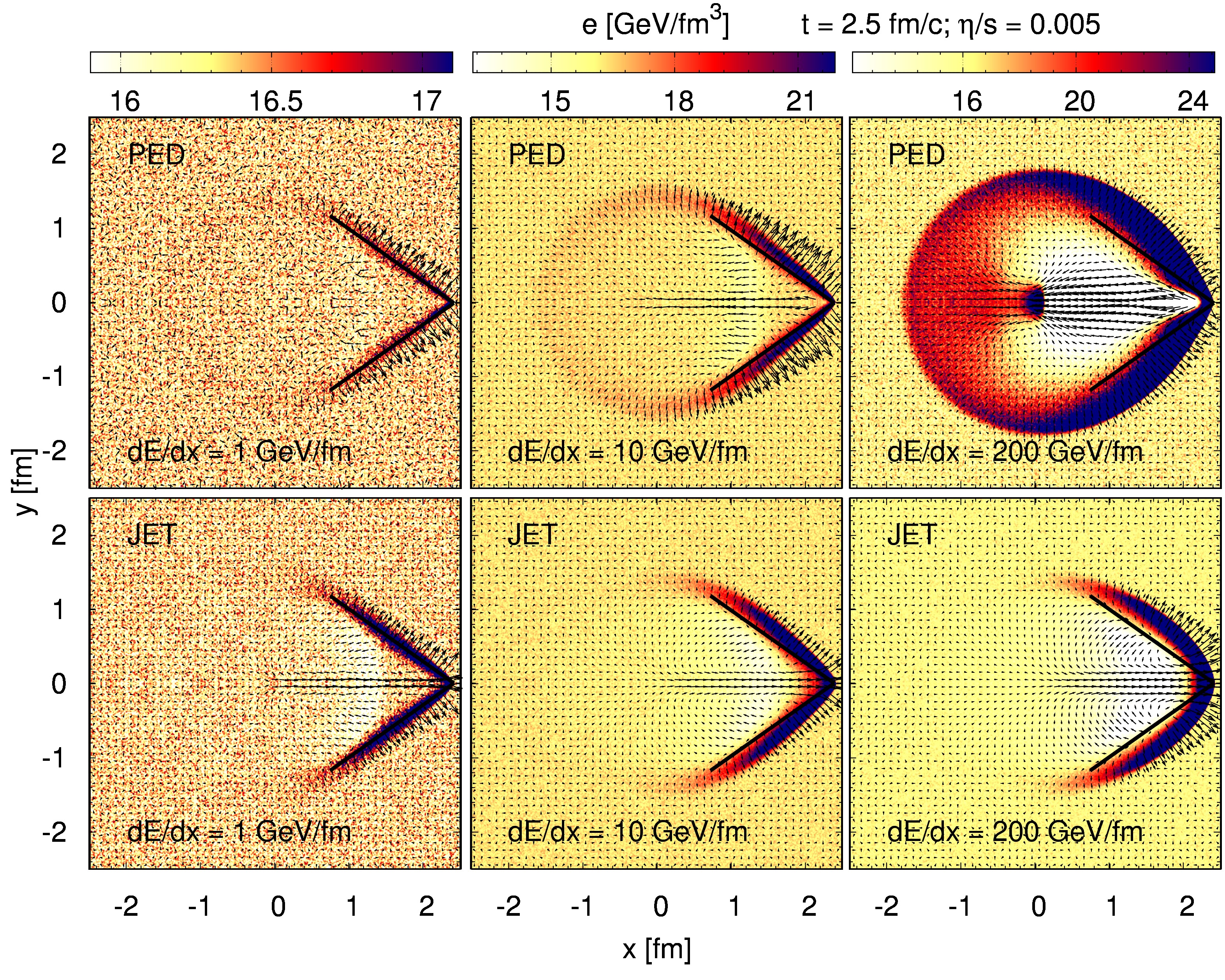}
\caption{(Color online) Shape of a Mach cone in the nearly ideal limit ($\eta/s = 0.005$)
shown for different jet scenarios and different energy deposition rates into
the medium, $dE/dx = 1, 10$ and $200$ GeV/fm. The upper panel shows the
pure energy deposition scenario (PED); the lower
panel shows the propagation of a highly energetic jet (JET) depositing
energy and momentum in $x$-direction. Depicted are the LRF energy density
within a specific range; as an overlay we show the velocity profile with
a scaled arrow length. The results are a snapshot of the evolution at $t = 2.5$ fm/c. In
addition we show the analytical solution for the ideal Mach cone in the
very weak perturbation case with the emission angle $\alpha_{\rm w}$.}
\label{fig:eDensity_idealDifSourceTerms}
\end{figure*}

In Fig.~\ref{fig:eDensity_idealDifSourceTerms} we show
the results for the PED in the upper panel and JET in
the lower panel using three different energy deposition
rates into the medium, $dE/dx = 1, 10$ and $200$ GeV/fm,
in the nearly ideal limit, i.e. $\eta/s \approx 0.005$.
We note that in general the maximum (minimum) energy density
in the simulations is larger (smaller) than the maximum (minimum) of the
energy density scales in the figures. Also the plotted arrow length of
the velocity profile is scaled. Both modifications are done to enhance
the readability of the figures.

In both scenarios, PED and JET, we observe a conical structure, but with
obvious differences. In the PED case with the isotropic energy deposition, a
circle of perturbations propagating in backward direction is visible. 
This is missing in the JET scenario because of the strong momentum
deposition in $x$-direction. Another difference is that in the JET scenario
a clearly visible head shock, i.e. a shock wave in the front of the 
jet perpendicular to the direction of the jet, appears.
This in turn is missing in the PED scenario. Furthermore, there is
a clear difference in the behavior of the matter behind the Mach
cones. In the JET case, the projectile induces a diffusion wake, where
the matter is flowing in the direction of the projectile. Whereas in
the PED scenario an opposite behavior is observed, i.e. there is an
(anti-)diffusion wake where the matter behind the cone is flowing in 
the backward direction. These observations are in qualitative 
agreement with the results from ideal hydrodynamics and transport 
calculations \cite{Betz:2008ka,Molnar:2009kx}.

Additionally, every scenario is compared to the ideal Mach cone
with $\alpha_w$ for a very weak perturbation shown in
Fig.~\ref{fig:eDensity_idealDifSourceTerms}.
Both scenarios provide evidence that the energy deposition rate
of the source influences the Mach angle $\alpha$ of the wings
according to Eq.~\ref{eq:IdealMachAngle}.
In both cases the shock front is curved because near the projectile
the disturbance of the media is strongest and the shock front moves
faster than the speed of sound. Farther away from the projectile a
part of the energy of the shock front has already dissipated
into the medium and as a result the perturbation gets weaker and
approaches a weak perturbation propagating with the speed
of sound.

In the JET scenario the energy of the jet $E_{ \rm jet}$ is
$20$, $200$ and $20000$ GeV (starting from the left in
Fig.~\ref{fig:eDensity_idealDifSourceTerms}). For our
calculations in the nearly ideal limit the energy of the jet does
not play any significant role. The only parameter which matters
is the average energy deposition rate. We will mention in
Sec.~\ref{sec:Viscosity} how the value of the jet
energy $E_{ \rm jet}$ and finite viscosity in the medium changes
the pattern of the Mach Cone.

%
%

\begin{figure}[h]
\centering 
\includegraphics[width=\columnwidth]{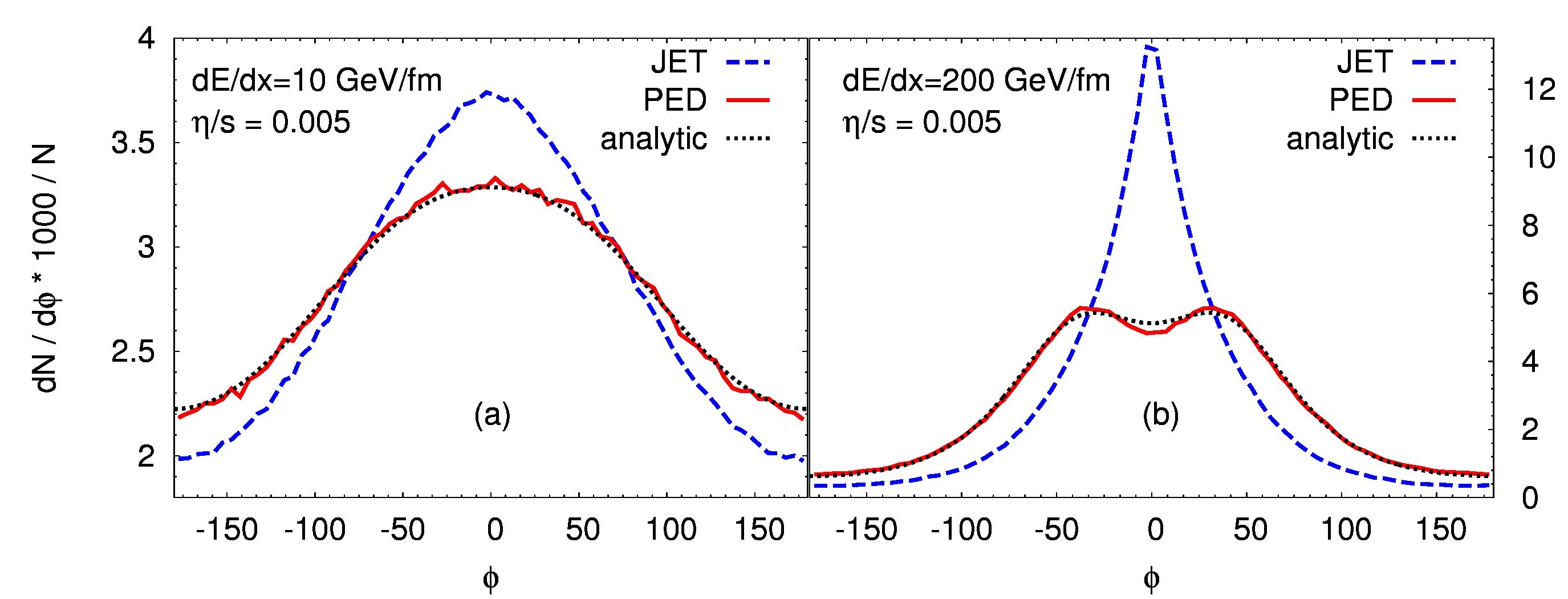}
\caption{(Color online) Two-particle correlations
$dN/(N d\phi)$ extracted from calculations shown in Fig.
\ref{fig:eDensity_idealDifSourceTerms}. The results are
extracted from calculations with $dE/dx = 10$ GeV/fm (a)
and $dE/dx = 200$ GeV/fm (b). Analytic solutions
extracted from Eq.~\eqref{eq:particleMultiCone} are shown
for $e_{ \rm cone}=22.15 \, \rm{GeV/fm^3}$ (a) and
$e_{ \rm cone}=62.55 \, \rm{GeV/fm^3}$ (b).
}
\label{fig:numTPC}
\end{figure}

We now want to address the question whether the Mach cone structures
observed in Fig.~\ref{fig:eDensity_idealDifSourceTerms} can be
regarded as the source of a double peak structure in two-particle
correlations. For this purpose we extract the particle distribution
$dN/(N d\phi)$ from BAMPS calculations. In Fig.~\ref{fig:numTPC}~(a)
we show the results for the energy deposition
rate $dE/dx = 10$ GeV/fm together with the analytical calculation using
Eq.~\eqref{eq:particleMultiCone}. To extract only the contribution
from the wings and to exclude of all other regions such as
(anti-)diffusion wake and back region (especially in the PED scenario),
a lower energy density cut at $20 \, \rm{GeV/fm^3}$ is applied.
Particles in cells with energy density lower than this value are
not considered in the extracted particle distribution (we note that
particles from the medium in rest automatically do not contribute
to the final profile). For the analytical solution taken from
Eq.~\eqref{eq:particleMultiCone} we use $e_{ \rm cone}=22.15 \, \rm{GeV/fm^3}$
and $v_{\rm coll} = 0.137$ ($e_{ \rm cone}$ represents the average
energy density on the Mach Cone wings extracted from the associated
numerical calculations). In both scenarios, PED and JET, as well
as in the analytical calculation we observe only a peak in the
direction of the source, but no double peak structure. This
finding is against all expectations from the naive picture of
a Mach cone. 

However, with a sufficiently higher energy deposition rate
the final picture changes significantly. In Fig.~\ref{fig:numTPC}~(b)
the results from BAMPS
calculations with $dE/dx = 200$ GeV/fm are shown. The lower
energy density cut is increased to $50 \, \rm{GeV/fm^3}$
because of the much higher energy deposition rate. For the analytic
calculation $e_{ \rm cone}=62.55 \, \rm{GeV/fm^3}$ with
$v_{\rm coll} = 0.537$ is selected. In the PED scenario,
as well as in the analytic model, the double peak structure
finally appears as long as the energy deposition rate and consequently
$v_{\rm coll}$ are sufficiently large. However, in the JET
scenario only a peak in the direction of the jet is visible.

We want to mention that in the PED scenario using special
momentum cuts (not considered in this work), i.e.
restricting the momentum integration in
Eq.~\eqref{eq:particleMultiCone} to a certain interval
one can always obtain a double peak in the distribution.
However, in the JET scenario a double peak never appears,
regardless of momentum cuts.

There are two main contributions to the structure of the 
two-particle correlation, one from the wings of the Mach
cone and one from the head shock region. The matter in
the wings is moving in the direction perpendicular to the
surface with some collective velocity $v_{\rm coll}$.
The larger the collective
velocity, the more strongly peaked are the local particle
distribution functions into this direction. From our simple
analytic model it is clear that the mere existence of the 
wings is not enough to have clearly visible peaks in the
correlation, but the local velocity of the matter
has to be sufficiently large. This is also confirmed
by the full simulations: If the energy deposition rate is
sufficiently large in the PED scenario, the double peaks
appear.

In principle, the same reasoning also works for the JET scenario.
However, in this case there is also a strong contribution 
from the head shock region, where the matter is moving
with large collective velocity. This collective motion is
in the direction of the projectile and results in a 
particle distribution function that is peaked in the same direction.
Although a double peak due to the Mach cone wings
still exists, the contribution of the head shock 
clearly dominates and overshadows the contribution from 
the wing regions (with spatial cuts to remove the head shock
the double peak appears again \cite{Bouras:2010nt}).
Thus, no double peaks appear in the JET scenario. 
%
%

\subsection{Effects of viscosity}
\label{sec:Viscosity}

\begin{figure*}[ht]
\centering 
\includegraphics[width=\textwidth]{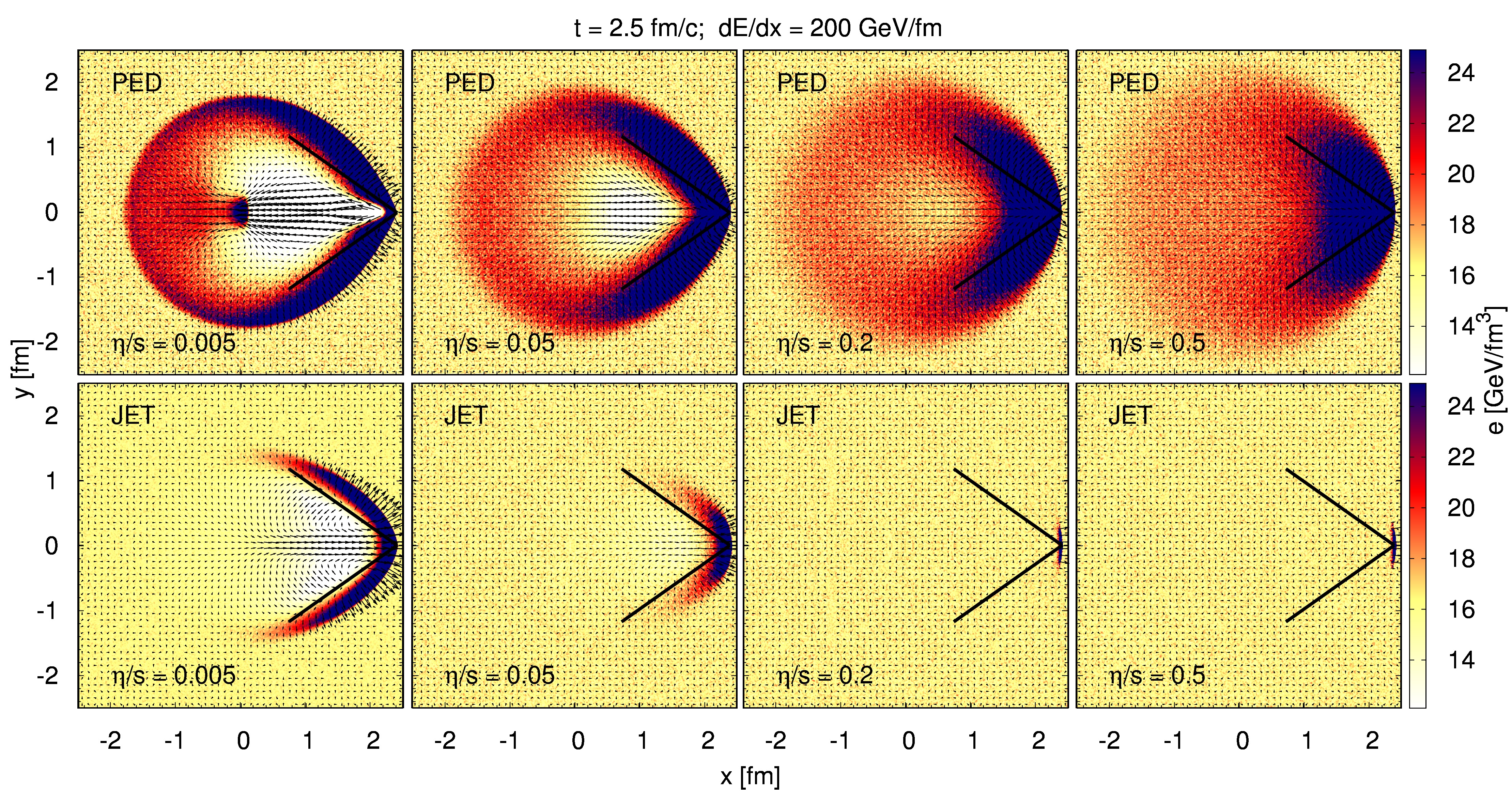}
\caption{(Color online) Transition from ideal to viscous Mach cones.
Shape of a Mach cone shown for different jet scenarios and different
viscosity over entropy density ratios, $\eta/s = 0.005$, $0.05$, $0.2$ and $0.5$.
The energy deposition is $dE/dx = 200$ GeV/fm. The upper
panel shows the pure energy deposition scenario (PED); the lower
panel shows the propagation of a highly energetic jet (JET) depositing
energy and momentum in $x$-direction. Depicted are the LRF energy density
within a specific range; as an overlay we show the velocity profile with
a scaled arrow length. The results are a snapshot of the
evolution at $t = 2.5$ fm/c. In addition we show the analytic solution
for the ideal Mach cone in the very weak perturbation case with the
emission angle $\alpha_w$.}
\label{fig:eDensity_viscDifSourceTerms}
\end{figure*}

In Fig.~\ref{fig:eDensity_viscDifSourceTerms} we show the
Mach Cone structure for both PED scenario (upper panel) and
JET scenario (lower panel) with $\eta/s = 0.005$, $0.05$,
$0.2$ and $0.5$ from left to right, respectively. The energy
deposition rate is fixed to $dE/dx = 200$ GeV/fm. In addition, 
$E_{ \rm jet} = 20000$ GeV is used in the JET scenario.
The chosen $\eta/s$ values are intended to cover the
nearly-ideal limit ($0.005$), the estimated QGP shear
viscosity over entropy density ratio
in heavy-ion collisions ($0.05$, $0.2$)\cite{Luzum:2008cw,Xu:2007jv}
and highly viscous limit where dissipative hydro calculations
are not reliable anymore ($0.5$) \cite{Bouras:2009nn}.

First, we note that if we observe the system at fixed time,
then in both scenarios the Mach cone structure smears out
and eventually disappears almost completely as the viscosity
increases. This is true for shock fronts as well as for the
(anti-) diffusion wake. The difference between the PED and
the JET case is that as $\eta/s$ increases, in the PED
scenario the resulting "Mach cone" solution covers
approximately the same spatial region regardless of a value
of $\eta/s$, while in the JET case the structure is
concentrated more and more near the projectile as the
viscosity increases. The reason for this is that in the
PED scenario the momentum from the projectile is
isotropically deposited into the medium, while in the
JET scenario the initial momentum dissipation is strongly
peaked into the direction of the projectile (the effect
in the JET scenario becomes even stronger with increasing
energy of the jet $E_{ \rm jet}$, since scattered the
particles are stronger forward-peaked). With a large
viscosity the re-scattering of the emitted particles from
the source is very rare. Thus, the larger the viscosity the
more the resulting solution reflects the details of the
projectile-matter interaction. 

We note that in both scenarios the projectiles are point-like
and initially the matter is homogeneously distributed.
Therefore, the only length scales that control the solution
are the mean free path, $\lambda_{\rm mfp} \propto \eta$,
and the energy deposition rate, ${\rm d}E/{\rm d}x$.
Thus, we expect a similar scaling behavior as in the
one-dimensional Riemann problem \cite{Bouras:2009nn}. For
example, the energy density profiles for two different
shear viscosities $\eta$ and $\eta'$ are related by
\begin{align}
&e(t-t_0, x-x_0, y-y_0, \frac{{\rm d}E}{{\rm d}x}, \eta) \\
&= e'(\frac{t-t_0}{C}, \frac{x-x_0}{C}, \frac{y-y_0}{C}, \frac{1}{C^{N-1}} \frac{{\rm d}E}{{\rm d}x}, \eta') \, .
\label{eq:scaling}
\end{align}
where the scaling factor $C = \eta'/\eta$, and $x_0$ and
$y_0$ are the coordinates of the projectile at the time $t_0$.
Here, $N$ counts the physically relevant number of dimensions
in space. In our case we have $N = 2$ since we keep the $z$-direction
as homogeneous.

Using this scaling behavior, we can interpret
Fig.~\ref{fig:eDensity_viscDifSourceTerms} as
a time-evolution of the solution, with a larger viscosity
corresponding to an earlier time and with an appropriate
scaling of the energy-deposition rate. For example, the solutions
with $\eta/s = 0.5$ in the right-most panel of
Fig.~\ref{fig:eDensity_viscDifSourceTerms} will evolve to
the ones with $\eta/s = 0.05$ at time $t = 25$ fm/c (with
the appropriate scaling of the $x-$ and $y-$axis and the energy deposition rate).
Although,
from Fig.~\ref{fig:eDensity_viscDifSourceTerms} the Mach
angle apparently changes with the viscosity, this is a
transient effect related to a finite formation time of
the Mach cone with non-zero viscosity. The viscosity affects
the width and formation time of the shock front, but not its
speed of propagation, i.e. the relation \eqref{eq:v_shock}
still holds for non-zero viscosity. Asymptotically, the Mach
cone angle will be the same regardless of the value of $\eta/s$.

\begin{figure}[h]
\centering 
\includegraphics[width=\columnwidth]{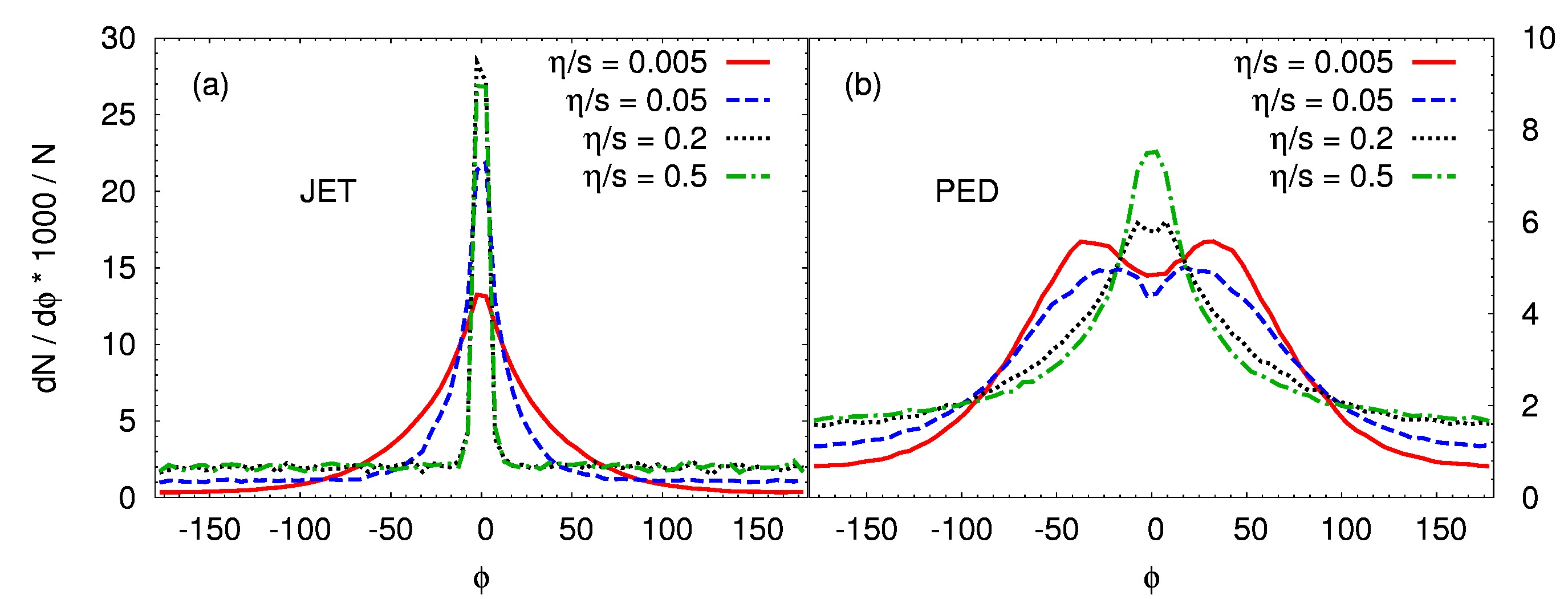}
\caption{(Color online) Two-particle correlations
$dN/(N d\phi)$ for different viscosities extracted from
calculations shown in Fig.~\ref{fig:eDensity_viscDifSourceTerms}.
The results are shown in the for the JET (a) and PED (b)
scenario for $dE/dx = 200$ GeV/fm.
}
\label{fig:num_visc_TPC}
\end{figure}

In Fig.~\ref{fig:num_visc_TPC} we show the two-particle correlations
for the solutions from Fig.~\ref{fig:eDensity_viscDifSourceTerms}.
The procedure is similar to the one
discussed for Fig.~\ref{fig:numTPC}. The lower energy density
cut is chosen to be $50 \, \rm{GeV/fm^3}$. For the JET scenario
(a), the peak in direction of the jet becomes
sharper with larger viscosity and no other appreciable effect
originating from viscosity is visible. In contrast, for the PED
scenario (b) the viscosity destroys the double
peak structure. If the viscosity is very large, only a peak in
direction of the jet is visible. As above, these results can
also be read as a time-evolution of the solution. Fig.~\ref{fig:num_visc_TPC}
shows how the angular distribution of the emitted particles
widens with time, or equivalently with increasing $\eta/s$.

%
\section{Conclusions}
\label{sec:conclusions}
%

In summary, we have investigated the structure of relativistic
Mach cones by using a microscopic transport model. The simulations
were realized by using two different types of sources propagating
through the matter, the PED and JET scenarios. The effect of the
strength of the projectile-matter interaction was studied by
varying the energy dissipation rate from the projectile to
the matter. Furthermore, the effect of the viscosity of the 
matter was investigated by adjusting the shear viscosity over
entropy density ratio $\eta/s$ from $0.005$ to $0.5$. 

We observed the conical structures form for both types of
sources in the nearly perfect fluid limit, similar to observations
in \cite{Betz:2008ka}, with the Mach Cone angle depending on the 
energy dissipation rate. We also demonstrated that the non-vanishing
viscosity tends to destroy the clear conical structure. 
By using a scaling of the solutions, we argued that increasing
the viscosity has the same effect as looking at the solution at 
an earlier time. The larger the viscosity or, equivalently, the 
less time the Mach cone has to develop, the more the structure 
of the solution depends on the details of the projectile-matter 
coupling.

Although Mach cone-like structures are observed in BAMPS
calculations for different energy and momentum deposition
scenarios they are not necessarily associated with double
peak structures in the azimuthal particle distributions in
$dN/(N d\phi)$. We found that only the PED scenario together
with a rather high energy deposition rate lead to a double
peak structure, which otherwise cannot be observed because
of the strong diffusion wake and head shock. However, the PED
scenario has no correspondence in heavy-ion physics. On the
other hand, the JET scenario is a simplified model but
nevertheless demonstrates that a double peak structure
cannot be produced by jets with energy and momentum deposition.
We expect that our conclusions will still be valid for realistic
jets and energy loss scenarios \cite{Fochler:2010wn}.
In addition, a clear Mach cone structure, which is necessary
but not in itself sufficient to produce a double peak structure
in two-particle correlations, will hardly develop in a system
of the size and finite viscosity relevant for HIC. We thus
conclude that the double peak structure is not the appropriate
observable for the signal of Mach cones in heavy-ion collision
experiments.

%
%
The authors are grateful to F. Reining, E. Molnar, B.\ Betz, J.\ Noronha,
G. \ Torrieri, P. \ Huovinen, J. Ulery, P. Rau and H.\ St\"ocker
for discussions and to the Center for Scientific Computing (CSC)
at Frankfurt University for the computing resources. I.\ Bouras is grateful
to HGS-Hire. The work of H.\ Niemi was supported by the Extreme
Matter Institute (EMMI). This work was supported by the Helmholtz
International Center for FAIR within the framework of the LOEWE
program launched by the State of Hesse.



\end{document}